\documentclass[10pt,letterpaper,twocolumn]{article} 

\usepackage{ol2}
\usepackage[draft]{hyperref}
\usepackage{amsmath}
\usepackage[utf8]{inputenc} 

\begin{document}

\twocolumn[ 

\title{Injection-locking of violet laser diodes with a 3.2GHz offset
frequency for driving Raman transitions in $^{43}$Ca$^+$}

\author{B. C. Keitch$^{1,*}$, N.R. Thomas, D. M. Lucas$^2$}
\address{
$^1$IQE, ETH Z\"urich, Schafmattstrasse 16, 8093 Zurich, Switzerland\\
$^2$Department of Physics, University of Oxford, Parks Road, Oxford OX1 3PU, U.K.\\
$^*$Corresponding author: bkeitch@ethz.ch
}

\begin{abstract}
Two cw single-mode violet (397nm) diode lasers are locked to a single
external-cavity master diode laser by optical injection locking. A
double-pass 1.6GHz acousto-optic modulator is used to provide a 3.2GHz offset frequency
between the two slave lasers. We achieve up to 20mW usable output in each slave
beam, with as little as 25$\mu$W of injection power at room temperature. An optical heterodyne
measurement of the beat note between the two slave beams gives a linewidth of
$\le$10Hz at 3.2GHz. We also estimate the free-running linewidth of the master
laser to be approximately 3MHz, by optical heterodyning with a similar device.\end{abstract}


 ] 

\section{Introduction}
\noindent
Injection-locking laser diodes has been a standard technique in the infrared
for some time \cite{Bouyer:96}, and has useful applications in spectroscopy, metrology and interferometry. It can also be used for driving Raman transitions, where a stable frequency offset is required, but the absolute wavelength is not critical \cite{Angel:95}.  Offsets can be added to the master beam using an AOM to
generate two beams with a tunable frequency difference. This makes the technique very useful for quantum information experiments
with trapped ions and neutral atoms \cite{McKay:11}.

The availability of blue and violet diode lasers is a relatively recent
development; however it
 has already been shown that it is possible to directly inject a bare
InGaN laser diode from an Extended Cavity Diode Laser (ECDL) \cite{Komori:03,Hirano:05}. 

We are interested in Raman transitions between two of the hyperfine states of the $4S_{1/2}$ ground level of a $^{43}$Ca$^+$ ion for use in a quantum information processing scheme\cite{Steane:06}. 
Two beams drive Raman transitions
from $4S_{1/2}$(F=4) to $4S_{1/2}$(F=3) 
 with a 
frequency difference of 3
225 608 286Hz \cite{Arbes:94}, via the
$4P_{1/2}$ level at a wavelength of 397nm. Using laser beams rather than microwaves 
 provides a
mechanism of imparting coherent motion to a calcium ion in a Paul trap. This
allows both cooling of the ion and preparation of the ion in a given motional state. 

Whilst the frequency difference needs to be very stable, we 
off-resonantly drive the transition \cite{Ozeri:07}, and therefore the absolute wavelength is not
critical. Grating-stabilisation gives an output power of about 50\% of the free-running diode's power. Use of EOMs or direct
modulation of the laser are possible; however, unlike an AOM, this would not
allow spatial filtering of the upper-sideband from the lower-sideband and
carrier. Direct modulation of either beam is inefficient, due to the low diffraction efficiency (25\%) and low damage threshold ($\approx$1\,mW at the necessary beam-diameter) of high-frequency AOMs. By using an injection scheme, both slave lasers can give maximum power with the
required frequency offset. 

Currently tapered amplifiers are not commercially available at the required atomic wavelengths, though there has been recent work in this direction \cite{Koda:10}. It is also possible to use frequency-doubled systems \cite{Hayasaka:02} but these require more experimental complexity, and are less desirable for experiments of long duration.

In section \ref{section:linewidth} the
linewidth of an unlocked ECDL is measured, to give reference data to compare against the
locked system. In section \ref{section:injection} direct injection of a diode laser gives a powerful slave beam to both compare with the modulated slave, and
to use in our application. In section \ref{section:offsetinject} we demonstrate offset injection-locking using the AOM. Due to the delicate nature of AOM
crystals that are efficient at this drive-frequency, the requirement for
successful injection at very low powers is even more stringent.

\section{Laser linewidth}
\label{section:linewidth}

\begin{figure}[h]
\centering
\includegraphics[width=7.5cm]{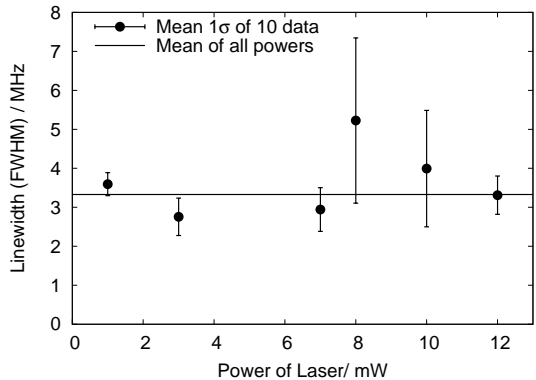}

\caption{Width (FWHM) of beat signal from two similar ECDLs. The signal was observed on
a spectrum analyser with a sweep time of 10ms. The width was found by fitting a
Lorentzian to the signal. This was repeated for different laser powers, where
both lasers were kept running within 0.2mW of each other. Each point is a mean
of ten readings with a 1$\sigma$ error-bar. The lasers were monitored for
single-mode behaviour and their absolute frequencies were monitored on a
wavemeter and kept about 3GHz apart.}

\label{linewidths}
\end{figure}

A heterodyning method (``beating'') was used to measure the linewidth of the
ECDL used in the following experiments. Two independent ECDLs were tuned to run
$\approx$3GHz apart, with similar powers and operating conditions. The InGaN diode was supplied by Nichia and used in a Toptica DL110 system. The ECDLs were
grating-stabilised in the Littrow configuration. Current control, temperature stabilisation and tuning of the
grating via a piezo-actuator were all included in the system. The temperature
stabilisation was specified to be 2mK RMS. The
grating was tuned to give wavelengths of 395.5nm, 396nm and 397nm, measured using a NIST LM10
wavemeter. In final use, the wavelength will be tuned near to the $4S_{1/2} \longleftrightarrow 4P_{1/2}$ transition in $^{43}$Ca$^+$ at
397nm.

The two beams were superimposed on each other using a beam-splitter, and
the resultant beam focused on to a high-speed photodiode (Newfocus 1437, 25GHz). The beat-note was amplified
and measured on a RF Spectrum Analyser (Agilent E4405B). The sweep time of the analyser
was reduced to a minimum of 10ms. However, the relative jitter of the two
free-running lasers during the sweep time gives a significant
contribution to the measured linewidth. The results are shown in figure
\ref{linewidths}. It can be seen that at a power of about 7mW the linewidth of
the ECDL is better than $3.2/\sqrt{2}=2.3$\,MHz. The specified linewidth of these ECDL systems is $<$5\,MHz over 100$\,\mu$s.

\section{Direct injection}
\label{section:injection}

\begin{figure}[h]
\centering
\includegraphics[width=7.5cm]{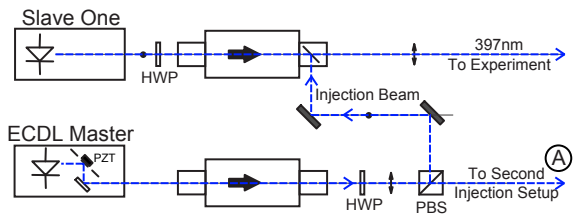}

\caption{Experimental setup showing the path of the injection light via the
Faraday isolators. Half-wave plates (HWP) and polarization beam splitting cubes (PBS) were used to control
power levels in each injection beam. Diagnostics (wavemeter and OSA) are not shown.}

\label{setup}
\end{figure}

An ECDL was used as a master laser to inject a bare diode laser (the slave). The
system is shown in figure \ref{setup}. The lasers were protected from stray back-reflection using
Faraday-rotator isolators (OFR IO-5-397-PHU-Z) which gave an isolation of
$>$30dB, with a transmission loss of about 1.5dB. The total power available
from the master ECDL (after its isolator) was 12\,mW, at a current of 51\,mA. The slave
laser had a power of about 20mW after its isolator, at maximum current. Both lasers were running close to room temperature.

Injection was investigated by launching a small amount of light from the master
into the slave via the side-port of the Faraday isolator. No beam-shaping was
done, as the master's beam shape closely matched that of the slave.

\begin{figure}[h]
\centering
\includegraphics[width=7.5cm]{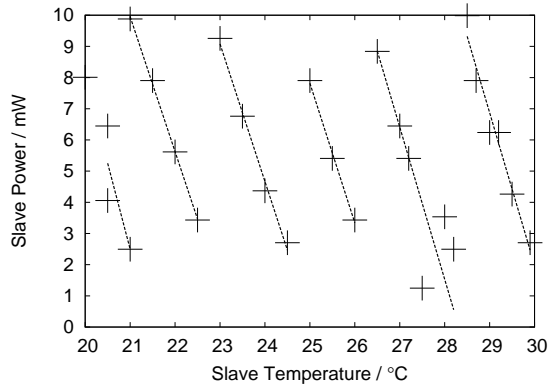}

\caption{Regions of stable injection in a laser diode, whilst injection-locked
to an ECDL.  Power is measured after the isolator. The lines are linear fits to the data points for each mode. Stable injection is possible anywhere along each line, but not in regions between the lines. Each point was taken by
allowing the slave temperature to settle, and then adjusting the current to obtain
stable, single-mode operation.}

\label{pvst}
\end{figure}

Throughout the experiment the output from each laser was monitored using an
optical spectrum analyser (OSA). The OSA was a CVI Technical Optics SA2-12, with
a free spectral range of 7.5GHz. Care was taken to ensure that the lasers were operating
single-mode. A plot of stable injection regions is shown in figure
\ref{pvst}. 

The power of the injection beam was reduced to the lowest level that stable
single-mode operation of the slave could be observed. Successful injection was achieved with an injection power entering the isolator
as low as 25$\mu$W. By monitoring both the
master and the slave laser on an OSA simultaneously, the slave could be seen to
``track'' the master as its wavelength was changed using a piezo
control on the grating. The master was not locked to a fixed reference, and so
instability in the master laser could cause the slave to lose its lock. However,
it was observed that the injection remained stable over a period of hours. 


\section{Injection with 3.2GHz offset}
\label{section:offsetinject}

\begin{figure}[h]
\centering
\includegraphics[width=7.5cm]{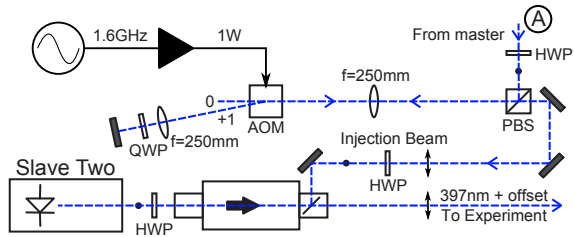}

\caption{Experimental setup showing the path of the injection light via an AOM running at 1.6GHz. The AOM is double passed
using a recollimating lens, a quarter-waveplate (QWP) and a mirror.}

\label{setup2}
\end{figure}

In the second part of the experiment, a fraction of the master power was passed
through an AOM using a double-pass setup. The setup is shown in figure \ref{setup2}, continuing from \ref{setup}. The AOM was a TeO$_{2}$ device (Brimrose TEF1600-150-395). Care was taken to keep the intensity below the damage
threshold (1W/mm$^2$). As there is evidence that high-powers rapidly age this
type of crystal \cite{Koerber}, the power was further reduced. The double-pass
arrangement allows the RF drive frequency to be changed without altering
alignment of the injection beam. The RF drive frequency was 1.6GHz, giving a frequency-offset of
3.2GHz after the double-pass. A lens
(f=250mm) was used to focus the beam into the AOM. The peak intensity at the
waist was 0.3W/mm$^2$ and efficiencies of 25\% on the first pass and 29\% on the second pass were
measured. An identical lens was used to recollimate the beam on the second pass
and it was found that the second pass was more efficient than the first due to the beam shape being better matched. Vertically polarised input light was the most efficient for the
AOM. After the first pass the polarisation was rotated returning through the
AOM and separated off using a PBS. The RF was generated using a synthesizer (Agilent E4422B) and amplified to 1W. 

\begin{figure}[ht]
\centering
\includegraphics[height=12em]{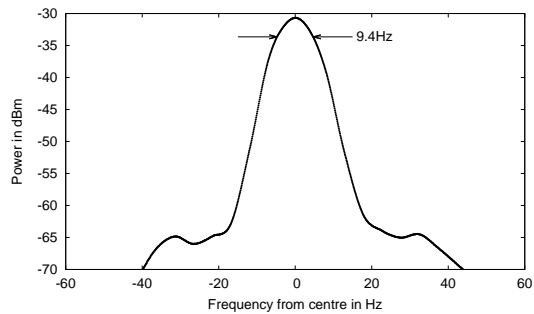}

\caption{The beat signal from two slave lasers both injection locked to the same
ECDL running at 397nm. The centre frequency is 3 223 997 775 Hz and the full width half-maximum is
9.4Hz. Video and resolution bandwidth are 10Hz.}

\label{beat}
\end{figure}

  Injection via the
double-pass AOM was achieved and the power was once again reduced to a minimum.
Successful injection at 60$\mu$W optical power at a frequency offset of 3.2GHz was
achieved. To measure the
relative stability (coherence length), 
 a heterodyne signal was generated as in section \ref{section:linewidth}. This beat
signal is shown in figure \ref{beat}.

\section{Conclusion}

Two violet diode lasers were injection locked to give two 20mW beams with a 3.2GHz frequency offset.
A heterodyne signal between the beams with a width of 10Hz at the $-$3dB points was observed, and
this was the minimum width that could be measured with a RF spectrum analyser that had a minimum bandwidth of 10Hz. This remained stable over a period of hours, without the master laser
being locked to any external frequency reference. The offset was tunable via the
RF source, and the injection remained aligned, due to the double-pass technique.
The absolute frequency was tunable using the external grating of the master
laser, along with its current and temperature.

This technique is suitable for high-fidelity experiments for quantum information processing and to drive coherent motion of a calcium ion in a Paul trap. It can also be extended to other Raman transitions at other wavelengths. By showing that successful injection can be obtained at very low power we avoid the complication of amplified or frequency-doubled systems and avoid damage to the AOM crystal.

We would like to thank Ursula Pavlish for her work in the laboratory.



\end{document}